\documentclass[a4paper,10pt]{article}
\usepackage[utf8]{inputenc}
\usepackage[left=2.5cm,right=2.5cm,top=1.5cm,bottom=2.5cm,footnotesep=0.5cm]{geometry}

\usepackage{hyperref}

\usepackage{amsmath}
\usepackage{amsfonts}
\usepackage{amssymb}
\usepackage{graphicx}
\usepackage{fancyhdr}

\usepackage{amsthm}
\theoremstyle{definition}
\usepackage{hyperref}

\usepackage{booktabs}

\usepackage{multirow}

\usepackage{xcolor}

\usepackage{subcaption}
\begin{document}
\fancyhf{}
\renewcommand{\headrulewidth}{0pt}
\renewcommand{\footrulewidth}{0.4pt}
\pagenumbering{gobble}
% \fancyfoot[C]{Submitted to BIAS 2023}
\fancyfoot[C]{BIAS 2023 Conference}

\Large
 \begin{center}
\textbf{Improving Model's Interpretability and Reliability using Biomarkers}
% User Study: Improving Model's Interpretability and Reliability using Biomarkers 
\hspace{15pt}

% Author names and affiliations
\large
% Wily E. Coyote \\
Gautam Rajendrakumar Gare$^1$, 
Tom Fox$^3$,
Beam Chansangavej$^3$
Amita Krishnan$^3$,
Ricardo Luis Rodriguez$^2$,
Bennett P deBoisblanc$^3$,
Deva Kannan Ramanan$^1$,
John Michael Galeotti$^1$
% John Michael Galeotti$^1$,

% Gautam Rajendrakumar Gare, Tom Fox, Beam Chansangavej, Amita Krishnan, Ricardo Luis Rodriguez, Bennett P deBoisblanc, Deva Kannan Ramanan, John Michael Galeotti

% \small Acme University of Explosive Research
% \small Robotics Institute, Carnegie Mellon University, USA Dept. of Pulmonary and Critical Care Medicine, Louisiana State University, USA Cosmeticsurg.net, LLC, Baltimore, USA\\
\small $^1$Carnegie Mellon University, $^2$Cosmeticsurg.net, $^3$Louisiana State University\\

\end{center}

\hspace{10pt}

\normalsize

% \vspace{-3em}

\begin{abstract}

    Accurate and interpretable diagnostic models are crucial in the safety-critical field of medicine. We investigate the interpretability of our proposed biomarker-based lung ultrasound diagnostic pipeline to enhance clinicians' diagnostic capabilities.
    The objective of this study is to assess whether explanations from a decision tree classifier, utilizing biomarkers, can improve users' ability to identify inaccurate model predictions compared to conventional saliency maps.
    Our findings demonstrate that decision tree explanations, based on clinically established biomarkers, can assist clinicians in detecting false positives, thus improving the reliability of diagnostic models in medicine.

\end{abstract}

\section{Introduction}
% Accurate diagnostic models are crucial for safety-critical applications, especially in medicine
% We propose to decouple feature learning from downstream lung ultrasound (LUS) tasks by introducing an auxiliary pre-task of visual biomarker classification. model as a bottleneck with an that enforces the models to go through a/
% The development of accurate and dependable diagnostic models is crucial for safety-critical applications, particularly in the field of medicine

% The development of accurate and dependable diagnostic models is crucial for safety-critical applications such as medicine. 
The development of accurate and dependable diagnostic models holds significant importance in the safety-critical field of medicine. Our research, described in \cite{Gare2022LearningTasks}, proposes a learning mechanism where the neural network model is enforced to go through an interpretable feature bottleneck of clinically established lung ultrasound biomarkers. As the feature encoder extracts biomarkers well-known to ultrasound radiologists, such expert users can now easily verify the black-box feature extractor output. The downstream classifiers (such as decision trees) that operate on these biomarkers are inherently interpretable as they operate on known features. Consequently, we have achieved a fully interpretable diagnostic model that adheres to the interpretability principles outlined in \cite{Lipton2016TheInterpretability}. In this work, we conduct a user study to formally assess the interpretability of our proposed biomarker model.
% In this work, we conduct a user study to formally assess the trustworthiness of our proposed biomarker model.
%%The development of accurate and dependable diagnostic models is critical in the field of medicine, where inaccurate model output can harm patients.

\begin{figure}[h]
% \begin{figure}
% \includegraphics[width=\linewidth, height=7cm]{Overview-interp.png}
\includegraphics[width=\linewidth, height=7cm]{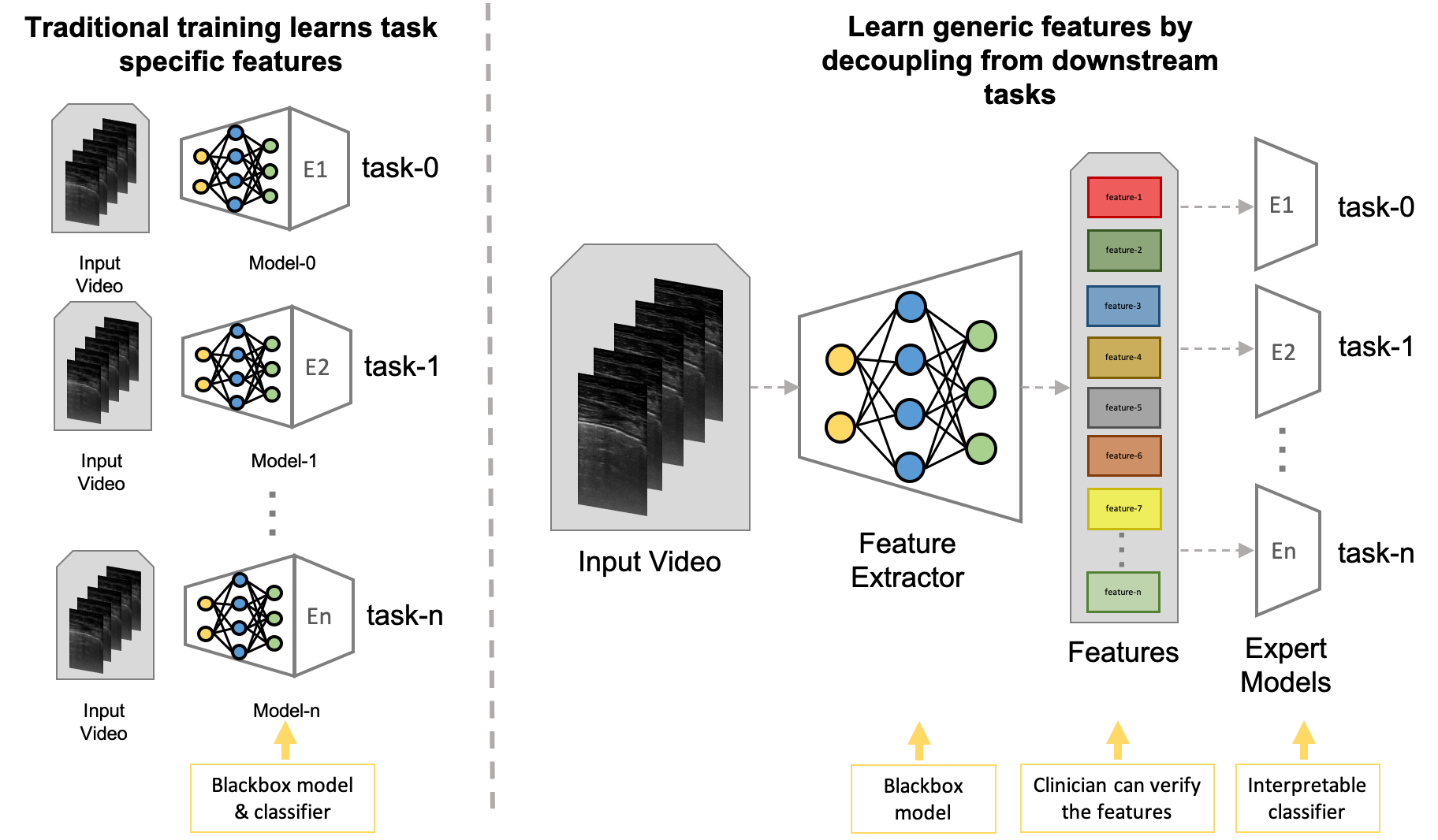}
\caption{
% In contrast to conventional approach that learn task-specific feature, we proposed in \cite{Gare2022LearningTasks} the decoupling of feature extraction from end-tasks by learning generic biomarker features.
% The biomarker model condenses the video into  multi-purpose biomarker feature attributes on which light weight task specific \emph{expert} models which are learnt with comparable performance.
% 
In contrast to conventional approach that learn task-specific feature, we proposed in \cite{Gare2022LearningTasks} the decoupling of feature extraction from end-tasks by enforcing models to go through an interpretable feature bottleneck of clinically established biomarkers. With help of clinicians we defined 38 lung ultrasound biomarkers that match/exceed DNN performance especially at low data regime. As the feature encoder extracts biomarkers well-known to ultrasound radiologists, clinician can now easily verify the black-box feature extractor output. Also, the downstream classifiers (such as decision trees) that operate on these biomarkers are diagnosable as they operate on known features. Thus giving rise to a fully interpretable diagnostic model.
} 
\label{fig:overview}
\end{figure}

% Lung ultrasound has proven to be a reliable diagnostic tool to clinicians especially since the COVID-19 pandemic \cite{Cogliati2021LungExperiences}. With this there has been a greater development of AI based approaches on lung ultrasound for diagnosing diseases. \cite{Roy2020DeepUltrasound1,Born2021AcceleratingAnalysis,GareTheAI,Xue2021ModalityInformation1,Gare2021DenseDetectionf}. With the proliferation on AI based diagnostic model, there is a need of reliable and interpretable techniques such that clinicians using these models can get a clearer understanding of these black-box based diagnostic models, inorder to ensure accurate and safe patient care in the safety critical. 

Lung ultrasound has emerged as a dependable diagnostic tool for clinicians, particularly in the context of the COVID-19 pandemic \cite{Cogliati2021LungExperiences}. This has spurred a rapid development of AI-based approaches for interpreting lung ultrasound images to aid in disease diagnosis \cite{Roy2020DeepUltrasound,Born2021AcceleratingAnalysis,GareTheAI,Xue2021ModalityInformation,Gare2021DenseDetectionf,Gare2022WeaklyUltrasound}. As the use of AI-based diagnostic models becomes more widespread, there is a pressing need for reliable and interpretable techniques. These techniques are essential to ensure that clinicians can gain a clear understanding of the workings of these "black-box" models. This understanding is crucial for accurate and safe patient care, particularly in the safety critical field of medicine.

The objective of this study is to evaluate whether explanations obtained from a decision tree classifier (built on our biomarkers \cite{Gare2022LearningTasks}) improves the ability of users to identify inaccurate model predictions. %occurring due to the model relying on false evidence.
Decision explanations depict the decision logic that the model used to derive its predictions. Saliency maps are a widely used post-hoc analysis tool that provides valuable information by highlighting the input image regions that influence the classifier's output \cite{SelvarajuGrad-CAM:Localization}. However, saliency maps cannot quantify the effect magnitude nor effect direction of the observed area of focus, limiting its utility in providing insight into the model's decision-making. The use of both saliency maps and decision trees may provide additional insights into the decision process and the evidence utilized by the model to make its predictions. Using this added insight, users may be able to more accurately determine when a model's output should be accepted and when it should be rejected. 
% Can't saliency

% saliency maps cannot quantify the effect magnitude of nor effect direction of the observed area of focus

% Saliency-maps depict the regions that the AI model focused on in the input clip. It is color-coded by temperature, with blue indicating low-attention and red indicating high-attention. Decision-explanations depict the decision logic that the model used to derive its predictions. It is an ordered set of rules, from high-to-low, that the model used based on the biomarkers it detected.

% Saliency-maps and decision-explanations can provide complementary information to saliency-maps. Saliency-maps: use color coding to show regions of the input image that contribute more or less to the model's prediction.
% In contrast, decision-explanations present the decision logic that the model used to arrive at its predictions as an ordered set of rules based on the biomarkers it detected, arranged in decreasing order of importance.

% \vspace{-1em}

\begin{figure}[h]
% \begin{figure}
\centering
\includegraphics[width=\linewidth]{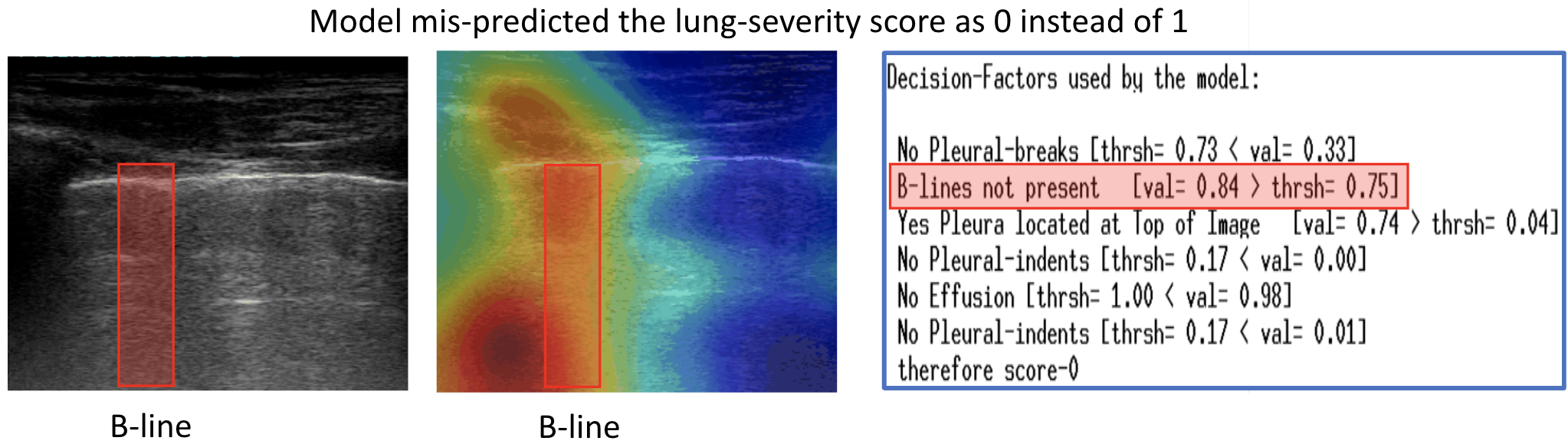}
\caption{
% Diagnoising the model's mis-prediction of lung-severity score as 0 instead of 1 for the above ultrasound clip, we can easily find by looking at the decision-tree explanations that the model ignored/missed the presence of B-line which led to the mis-prediction. Whereas relying solely on the saliency map, it is difficult to know the cause of mis-prediction as the models pays attention to the B-line as well as other regions across frames. Making it difficult to quantify the contribution of the B-line to the final prediction.
Diagnoising the model's mis-prediction of lung-severity score as 0 instead of 1 for the above ultrasound clip, by examining the decision-tree explanations, we can easily discern that the model overlooked the presence of B-lines, leading to the misprediction. In contrast, relying solely on the saliency map makes it challenging to identify the cause of the misprediction, as the model's attention is distributed across various regions in the frames, including the B-line. Consequently, making it difficult to quantify the contribution of the B-line to the final prediction.
} 
\label{fig:user-study-miss-pred}
\end{figure}
% \end{figure}

\section{Methods}
We conduct the study on 30 lung ultrasound (LUS) patient videos from the test dataset of \cite{Gare2022LearningTasks}. Of these videos, the model accurately predicted the lung-severity score (0, 1, 2, or 3) in 18 cases (60\% accuracy). Using the MLP classifier, we generated Grad-CAM \cite{SelvarajuGrad-CAM:Localization} based saliency maps, and we extracted the decision path of the videos from the decision tree classifier. Both the MLP and the decision tree classifiers produced identical predictions for all videos.

For each video clip, the clinicians were presented with the model's predicted lung severity score and were asked to evaluate its correctness with the assistance of either: saliency map, decision tree explanation, both, or none. The presentation of the videos and their variants were randomized to ensure that each evaluation was independent. The study was conducted with three LUS clinicians.

% We select 30 LUS videos obtained from 30 patients from the test dataset of \cite{Gare2022LearningTasks}, of which 18 videos the model's prediction was correct and the remaining 12 videos where the model's prediction was wrong. For all the videos the MLP classifier and the decision tree classifier had the same prediction. We generate GRAD-CAM \cite{SelvarajuGrad-CAM:Localization} based saliency maps using the MLP classifier and we extract the decision-path of the videos from the decision tree classifier.      

% For every clip, the clinician was presented with the model's predicted lung-severity score (0,1,2,3) \cite{GareTheAI} along with either the saliency-map, decision-path, or both or neither. The videos and the 4 variants were randomized such that the user did not see the same clip subsequently to ensure that every evaluation was independent. We conduct the study on three LUS clinician.

\begin{figure}[htb]
\includegraphics[width=\textwidth]{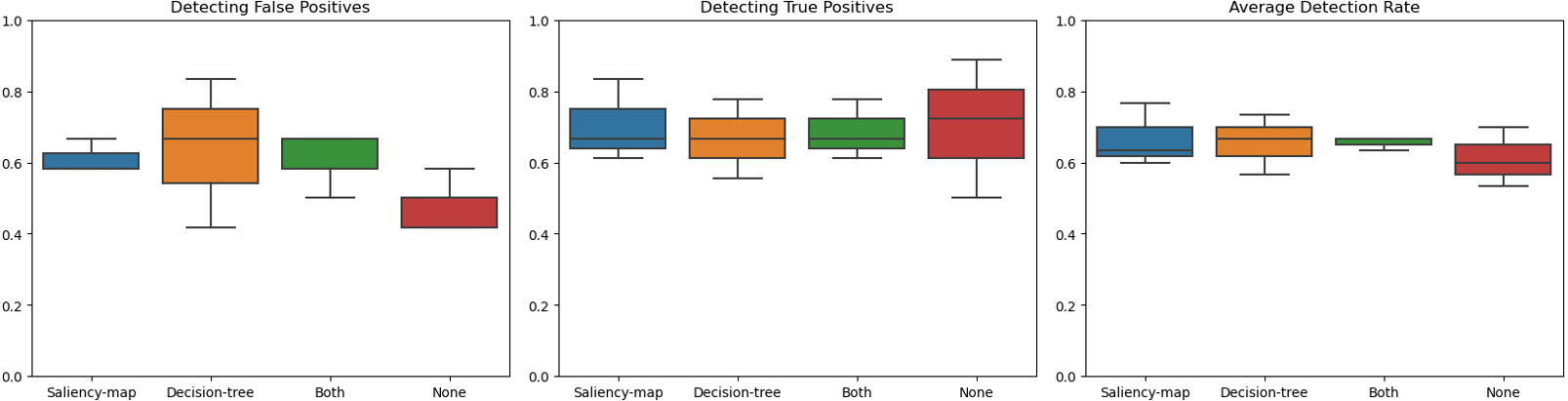}
% \caption{The checkbox annotator used for labeling the 38 biomarkers.} 
\caption{
Box plot showing the clinician's detection rates of the correctness of AI's output. We see that decision trees are effective at helping users detect false positive predictions. Overall we see that it's beneficial to have a model analysis tool to improve over baseline accuracy of 60\%, with saliency maps being the most beneficial.  
} 
\label{fig:user_study_results}
%\vspace{-0.8em}
\end{figure}

% a) Heat-maps

% b) Decision-explanations

% c) Both heat-maps and decision-explanations 

% d) None - just the input image and the model's prediction

% The clinicians were asked to evaluate if the model's predictions are based on valid-evidence and answer (4) questions regarding trust and confidence on the model's prediction. Finally, after going through all 30 clips, they answered which of the two (heat-maps or decision-explanations) was most useful. 

% \vspace{-2em}

\section{Results}
Figure~\ref{fig:user_study_results} shows that saliency maps are good at detecting true positives, whereas the decision tree is good at detecting false positives. Presenting both results in having the lowest standard deviation, indicating greater consensus among the clinicians. Finally, after going through all 30 clips, they rated the usefulness of the decision trees as 0.51 and heat-maps as 0.02 (average score on a scale of -1 to 1) and two clinicians favored having both while the third preferred decision trees over saliency maps.
% when asked how they were asked to indicate which of the two variants (saliency-maps or decision-explanations) they found to be more useful. 0.5105263158	0.02105263158

% Figure~\ref{fig:user_study_results} shows the number of times the users correctly identified when the model's prediction was correct (True-positive and True-negative cases) and when the model's prediction was wrong (False-positive and False-negative cases) for a particular case. The results indicate that heat-maps (without decision-tree), are good at detecting when model prediction is correct, whereas the decision-tree (without heat-maps), is good at detecting when model's prediction is wrong. However, overall heat-maps alone fared better than the decision-tree alone. Saliency-maps with decision-tree have the lowest standard deviation, indicating greater consensus among the users.

\begin{figure}[h]
\centering
\includegraphics[width=0.7\linewidth]{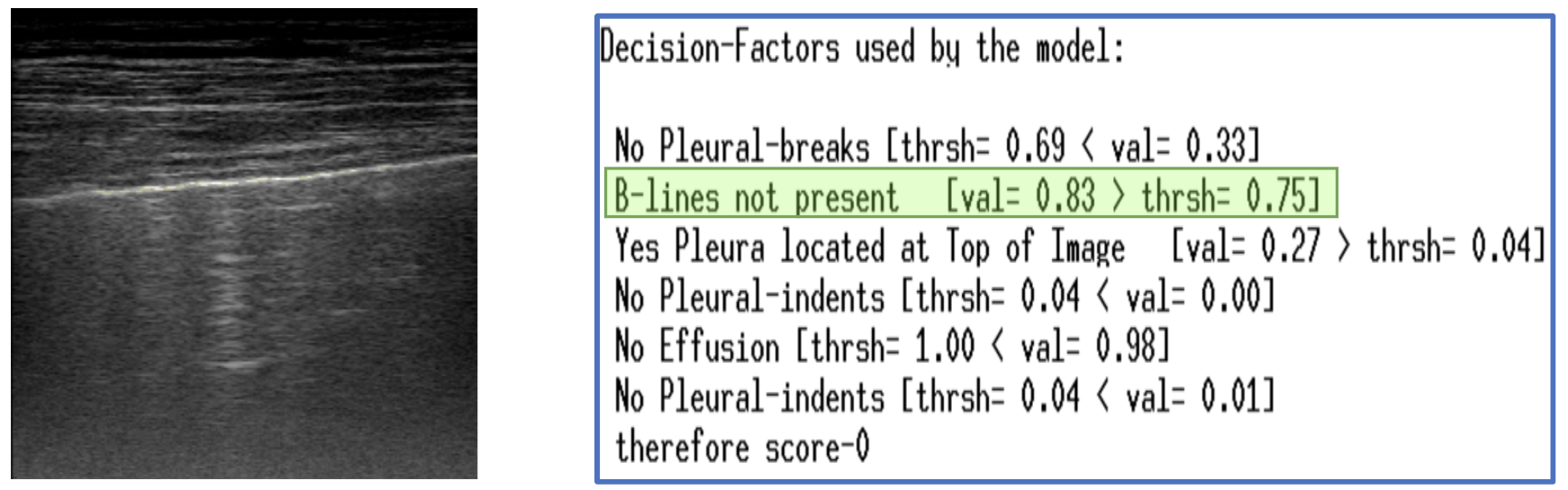}
\caption{
With our decision-tree explanations we can easily interpret the video. Which would be helpful for novice clinicians, who in this case could mistake the vertically stacked A-line bands for B-lines, whereas our decision tree explanation clearly shows no B-lines are present.
} 
\label{fig:user-study-corr-pred}
\end{figure}

\vspace{1cm}

\textbf{Discussion:} 
We find that interpreting ultrasound videos is more straightforward using our decision-tree explanations. This feature is particularly beneficial for novice clinicians. For example, in the ultrasound clip shown in Figure~\ref{fig:user-study-corr-pred}, a novice clinician might mistakenly identify the vertically stacked A-line bands as B-lines. However, our decision tree explanation clearly indicates the absence of B-lines.

In contrast, using saliency maps for interpreting ultrasound videos could be misleading. For example, in the ultrasound clip depicted in Figure~\ref{fig:user-study-miss-pred}, the model mispredicted the lung severity score as 0 instead of 1. Relying solely on the saliency map to diagnose this misprediction makes it challenging to identify the cause, as the model's attention is distributed across various regions in the frames, including the B-line. Consequently, quantifying the contribution of the B-line to the final prediction becomes difficult. However, with the decision-tree explanations, we can easily determine that the model overlooked the presence of B-lines, which led to the misprediction. 

% We find that it is easier to interpret ultrasound videos using our decision-tree explanations. This will be helpful especially for novice clinicians, consider the ultrasound clip in Figure~\ref{fig:user-study-corr-pred} where a novice clinician could mistake the vertically stacked A-line bands for B-lines, whereas our decision tree explanation clearly shows no B-lines are present.

% Diagnoising the model's mis-prediction of lung-severity score as 0 instead of 1 for the above ultrasound clip, we can easily find by looking at the decision-tree explanations that the model ignored/missed the presence of B-line which led to the mis-prediction. Whereas relying solely on the saliency map, it is difficult to know the cause of mis-prediction as the models pays attention to the B-line as well as other regions across frames. So we can't quantify how much the B-line is contributing to the final prediction.

% Deva: I do think it would be helpful to have a 1-2 sentence discussion of “what prevents people from ignoring the model prediction and just labeling the example themselves?” - e.g., it is expensive to label such data and user self-report that they prefer having the model output available to them.
% \textbf{Discussion:} 
% We see that access to analysis tools helps improve the reliability of models, where decision tree classifiers built on top of our biomarkers can help detect false positives, vital for diagnostic medicine. 

% However, overall saliency maps alone fared better than the decision tree. 

\section{Conclusion}
We conducted a user study comparing the interpretability of our proposed biomarker based lung ultrasound diagnostic pipeline. We compared decision tree explanations built on top of biomarkers with conventional saliency maps to help clinicians detect model mis-predictions.
We see that access to analysis tools helps improve the reliability of models, where decision tree classifiers built on top of our biomarkers can help detect false positives, vital for diagnostic medicine.

\vspace{1cm}
% \section{Acknowledgments}
\textbf{Acknowledgments:}
% \newline
% This research is supported in part by the Center for Machine Learning and Health (CMLH) Translational Fellowship, Carnegie Mellon University. 
% 
This research is supported in part by the Center for Machine Learning and Health (CMLH) Translational Fellowship, Carnegie Mellon University (CMU). The authors thank the clinicians at Louisiana State University (LSU) for their assistance with data collection and our CMU collaborators for their valuable insights. All data used in this study were collected under IRB protocol number 1509, titled Artificial Intelligence Interpretation of Lung Ultrasound Images, and were de-identified prior to transfer to CMU.

\newpage

{
% \small
% \bibliographystyle{abbrv}
\bibliographystyle{splncs04}
% \bibliography{refs}
\bibliography{references}
}

\end{document}